\documentclass[twocolumns]{aastex6}
\usepackage{graphicx}
\usepackage{amsmath}
\usepackage{longtable}
\usepackage{color}
\received{}
\revised{}
\accepted{}

\slugcomment{draft version March 28, 2018}
\shorttitle{BH-NS merger-powered short GRBs}
\shortauthors{Mu et al.}

\def\Swift{\textit{Swift}}
\def\s{\rm{s}}
\def\ej{\rm{ej}}
\def\keV{\rm{keV}}
\def\d{\rm{d}}
\def\r{\rm{r}}
\def\s{\rm{s}}
\def\e{\rm{e}}
\def\p{\rm{p}}
\def\erg{\rm{erg}}
\def\dec{\rm{dec}}

\begin{document}

\title{Central Engine-Powered Bright X-ray Flares in Short Gamma-Ray Bursts:\\
A Hint of Black Hole-Neutron Star Merger?}

\author{Hui-Jun Mu\altaffilmark{1}, Wei-Min Gu\altaffilmark{1},
Jirong Mao\altaffilmark{2,3,4}, Shu-Jin Hou\altaffilmark{5},
Da-Bin Lin\altaffilmark{6}, and Tong Liu\altaffilmark{1} }

\altaffiltext{1}{Department of Astronomy, Xiamen University, Xiamen,
Fujian 361005, China}
\altaffiltext{2}{Yunnan Observatories, Chinese Academy of Sciences,
650011 Kunming, Yunnan Province, China}
\altaffiltext{3}{Center for Astronomical Mega-Science, Chinese
Academy of Sciences, 20A Datun Road, Chaoyang District, Beijing, 100012, China}
\altaffiltext{4}{Key Laboratory for the Structure and Evolution
of Celestial Objects, Chinese Academy of Sciences, 650011 Kunming, China}
\altaffiltext{5}{College of Physics and Electronic Engineering,
Nanyang Normal University, Nanyang, Henan 473061, China}
\altaffiltext{6}{Guangxi Key Laboratory for Relativistic Astrophysics,
Department of Physics, Guangxi University, Nanning 530004, China}

\email{guwm@xmu.edu.cn}

\begin{abstract}

Short gamma-ray bursts may originate from the merger of double neutron
stars (NS) or that of a black hole (BH) and an NS.
We propose that the bright X-ray flare related to the central engine
reactivity may hint a BH-NS merger, since such a merger can provide
more fall-back materials and therefore a more massive accretion disk
than the NS-NS merger. Based on the observed 49 short bursts with
\Swift/X-ray Telescope follow-up observations,
we find that three bursts have bright X-ray flares,
among which three flares from two bursts are probably related to the central
engine reactivity. We argue that these two bursts may originate from the
BH-NS merger rather than the NS-NS merger.
Our suggested link between the central engine-powered bright
X-ray flare and the BH-NS merger event can be checked by the future
gravitational wave detections from advanced LIGO and Virgo.

\end{abstract}

\keywords{accretion, accretion disks --- gamma-ray burst: general
--- X-rays: general}

\section{Introduction}\label{sec1}

Gamma-ray bursts (GRBs) are generally divided into two classes
in the $T_{90}$ (duration) - HR (spectral hardness) space:
long-soft GRBs (LGRBs) and short-hard GRBs (SGRBs)
\citep{Kouveliotou1993},
or physically distinct types: Type I and II GRBs \citep{Zhang2009}.
It is well-known that LGRBs are associated with type Ib/c
supernovae (SNe), which are generally triggered by the collapse
of massive stars \citep[for a review, see][]{Woosley2006}.
SGRBs originate from the merger of double neutron stars (NS) or
that of a black hole (BH) and an NS
\citep[e.g.,][]{Paczynski1986,Paczynski1991,Eichler1989,Narayan1992}.

The compact binary mergers (NS-NS, BH-NS, BH-BH)
are the promising sources of strong gravitational waves (GWs).
Before the discovery of the GW event from the NS-NS merger,
the binary NS system had been detected in our Galaxy through
electromagnetic (EM) radiation \citep[e.g.,][]{Tauris2017}.
After the first detection of high-frequency GW signals from
the BH-BH merger \citep[e.g.,][]{Abbott2016a,Abbott2016b},
a new era of astronomy has been launched.
However, since the merger of BH-BH is not expected to produce
EM signals, a wide range of EM counterparts are expected for
the NS-NS and BH-NS mergers
\citep{Metzger2012,Zhang2016,Liu2017}.
On August 17, 2017, the Laser Interferometer Gravitational-Wave
Observatory (LIGO) and the Virgo detector simultaneously
detected a transient GW event of an NS-NS merger
\citep[GW 170817;][]{Abbott2017}.
SGRB 170817 was independently detected with a delay of about
$1.7 \rm{s}$ by both \emph{Fermi} Gamma Ray Burst Monitor
\citep{Goldstein2017,Zhang2018} and \emph{INTEGRAL} \citep{Savchenko2017}.
This is the first specific GW signals from the NS-NS merger,
and with an EM counterpart for the first time.

The ejection of neutron-rich matter is another key prediction
of the compact binary merger. The r-process-enriched ejecta
\citep[the so-called ``mini-SN", ``kilonovae"
or ``macronova"; see e.g.,][]
{Li1998,Metzger2010,Barnes2013,Yu2013,Metzger2017,Kasen2017}
\footnote{Recently, by testing the BH hyperaccretion
\citep[for a review, see][]{Liu2017}
inflow-outflow model for powering~GRBs, \citet{Liu2018}
and \citet{Song2017} proposed that another type of ``nova"
after the merger event may be detected.
In addition, \citet{Ma2018} proposed a
Blandford-Payne mechanism-powered merger-nova
in the BH hyperaccreting.}
from NS-NS or BH-NS mergers was widely studied \citep[e.g.,][]{Jin2016}.
On the other hand, the simulations of NS-NS mergers
\citep[e.g.,][]{Dietrich2015} and BH-NS mergers
\citep[e.g.,][]{Siegel2017}
showed an upper limit of the remnant disk mass,
i.e, $\sim 0.3M_{\sun}$.
The BH-NS merger simulations proposed that
the typical ejecta velocity is similar to the NS-NS case.
However, the ejecta mass $M_{\ej}$ generally increases with larger
asymmetry in the mass ratio, which can reach $0.1 M_{\sun}$
in the BH-NS merger \citep{Kawaguchi2015,Kawaguchi2016}.
According to the BH-NS merger model presented in
\citet{Tanaka2014} and \citet{Kawaguchi2016},
the ejecta mass can be estimated, i.e.,
GRB 050729 (I-band data, $M_{\ej} \sim 0.05 M_{\sun}$),
GRB 060614 (I-band data, $M_{\ej} \sim 0.1 M_{\sun}$), and
GRB 130603B (H-band data, $M_{\ej} \sim 0.03 M_{\sun}$).
The massive ejecta ($\ga 0.05 M_{\sun}$)
suggests that the progenitor of GRB 050709 is the BH-NS merger
\citep{Jin2016}, which is similar to the macronova
candidate of GRB 060614 \citep{Yang2015,Jin2015}.

In this work, we propose that the bright X-ray flares after the
prompt gamma-ray emission can also be a good probe for
distinguishing BH-NS and NS-NS mergers.
Several ideas have been proposed to explain the episodic phenomenon
of X-ray flares, i.e., related to the internal dissipation
\citep[e.g.,][]{Perna2006,Dai2006,Rosswog2007,Liu2017},
or to the external shock \citep[e.g.,][]{Proga2006,Giannios2006}.
The BH-NS merger simulations proposed that the typical ejecta
mass can reach $0.1 M_{\sun}$ \citep{Kawaguchi2015,Kawaguchi2016}.
\citet{Rosswog2007} shows a larger spread in the fallback behavior
in the BH-NS system than in the NS-NS one.
For an SGRB having bright X-ray flare with internal origin,
a more massive disk is required, corresponding to the BH-NS merger.
Our suggested link between the central engine-powered bright X-ray flare and
the BH-NS merger can be checked by the GW detectors.
The remainder is organized as follows.
Our data analysis and physical origin of the X-ray flares
are presented in Section~\ref{sec2}.
The accreted mass for X-ray flares is investigated in Section~\ref{sec3}.
A comparison of NS-NS and BH-NS mergers is studied in Section~\ref{sec4}.
Conclusions and discussion are presented in Section~~\ref{sec5}.

\section{data analysis and physical origin}\label{sec2}

\begin{figure*}
\centering
\includegraphics[angle=0,width=0.5\textwidth,height=0.35\textheight]{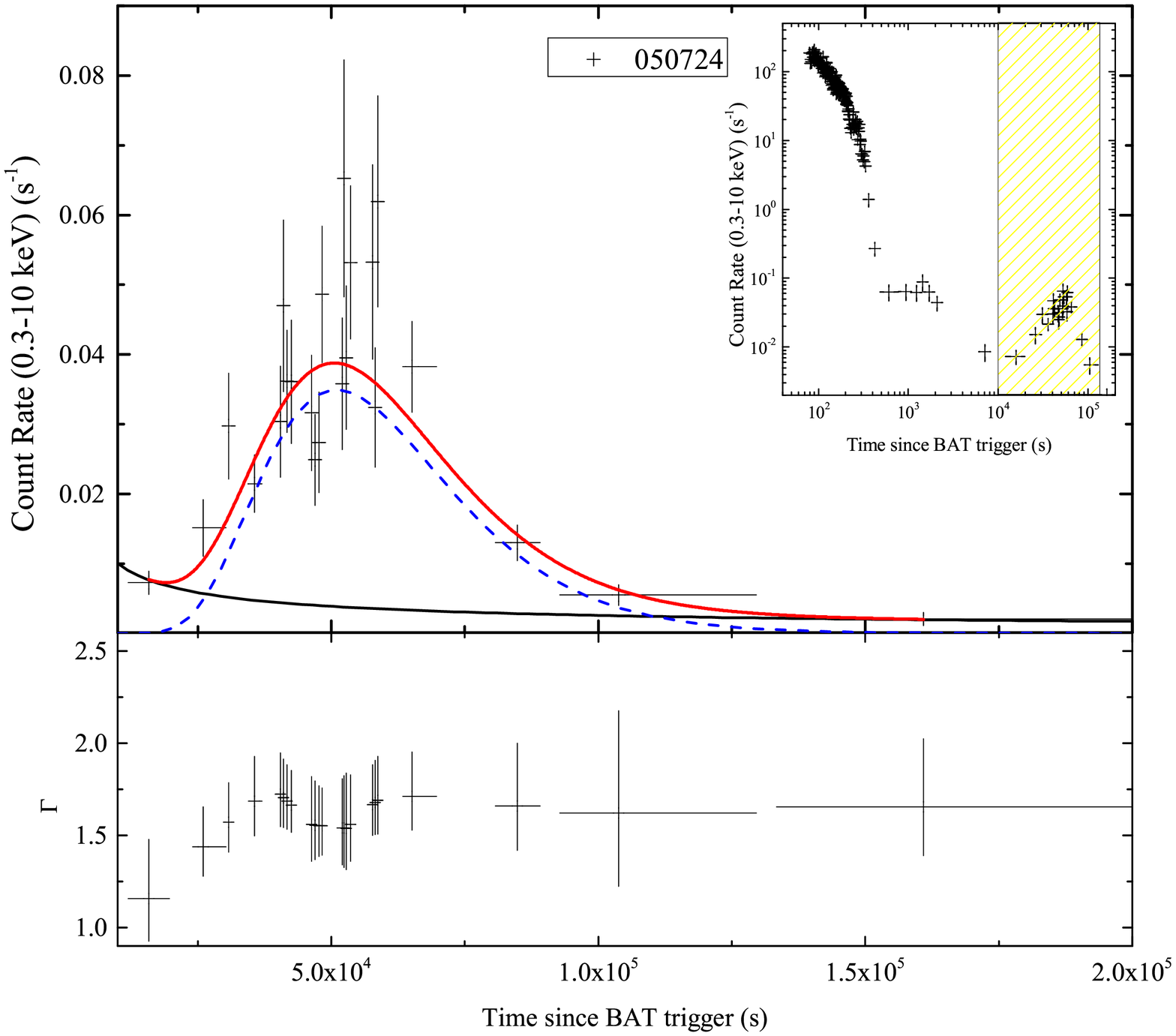}%
\includegraphics[angle=0,width=0.5\textwidth,height=0.35\textheight]{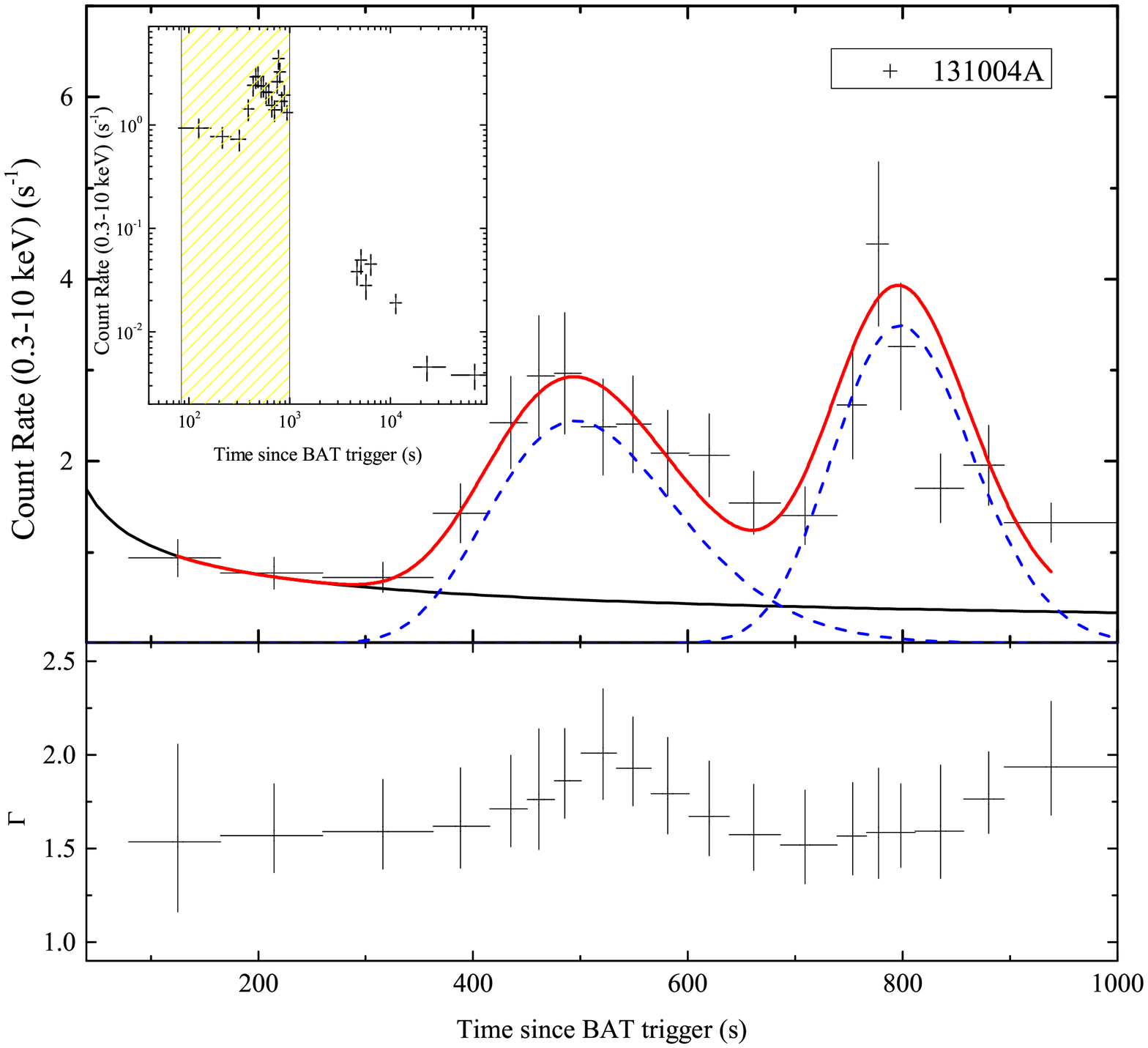}
\includegraphics[angle=0,width=0.5\textwidth,height=0.35\textheight]{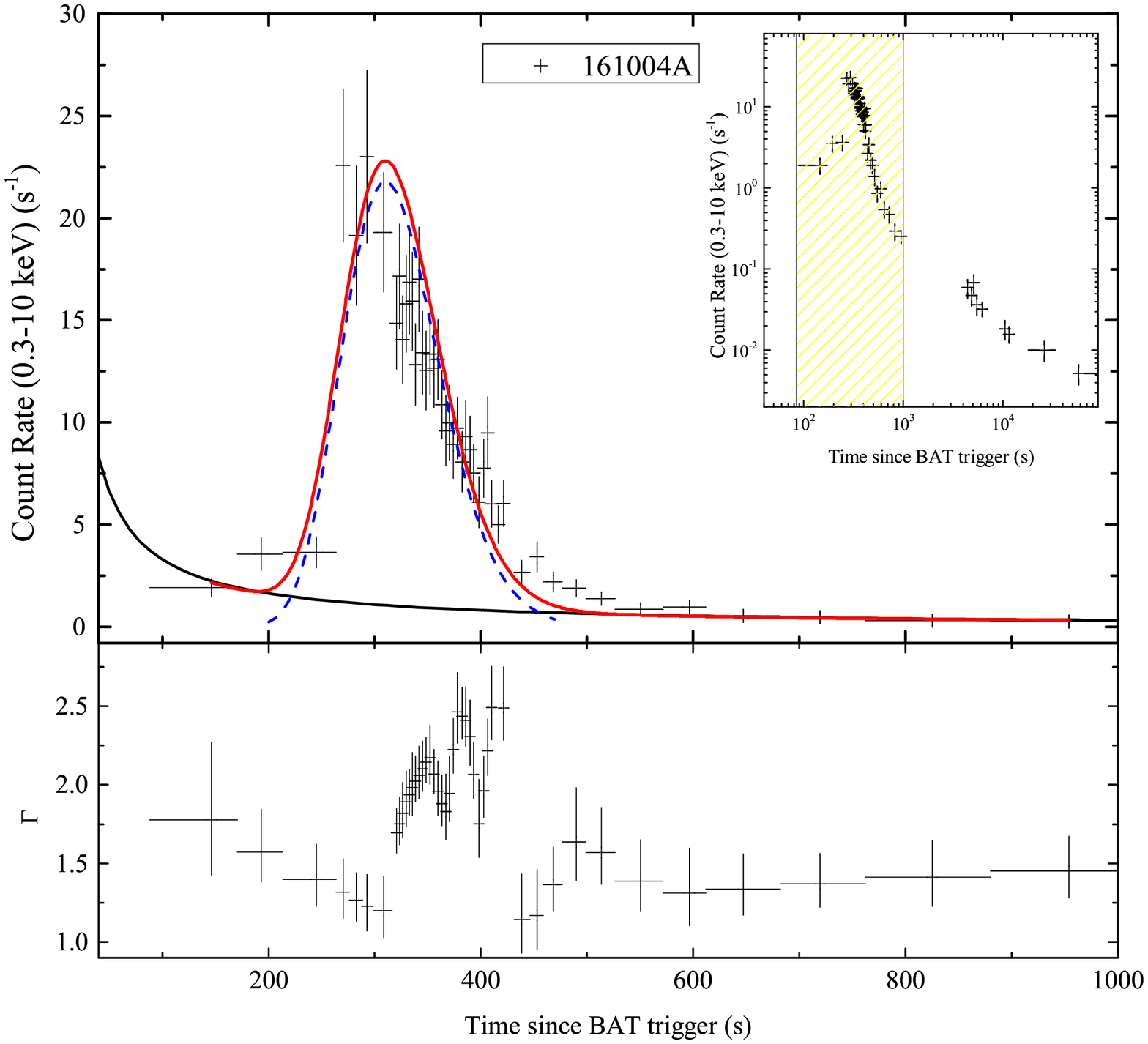}
\caption{
Upper panel: 0.3-10 \keV count-rate light curve of the three SGRBs in our sample.
Red solid line: best fitting for the total emission;
blue dashed line: best fitting for the bright X-ray flare;
black solid line: the smooth continuum contribution shown by a simple power-law.
Insert: complete \textit{Swift}/XRT light curve.
The yellow area shows the total emission selected in our fitting.
Lower panel: spectral photon index evolution.
}
\label{F1}
\end{figure*}

\begin{figure*}
\centering
\plotone{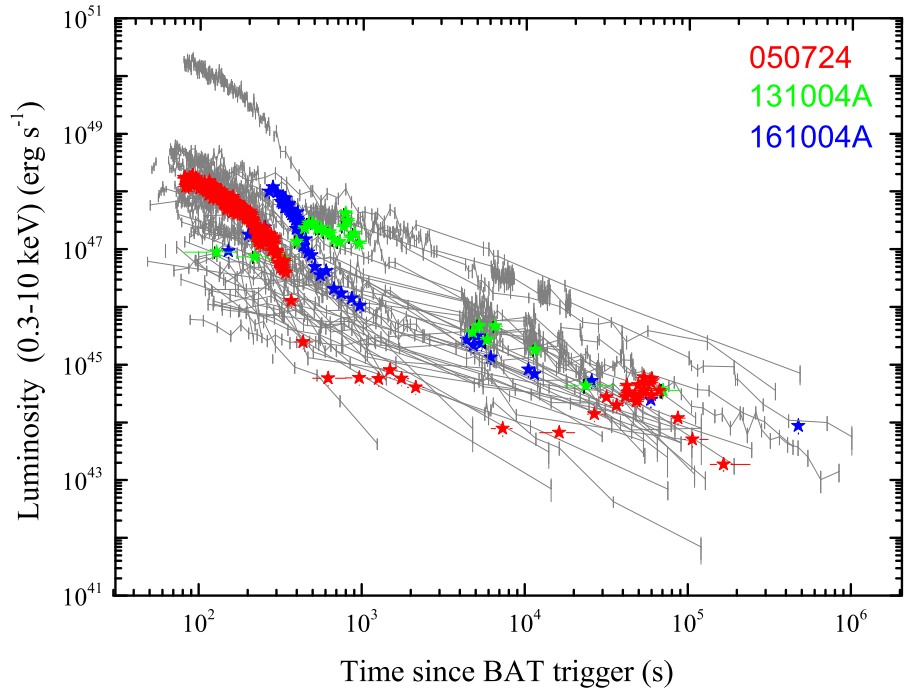}
\caption{
All the X-ray light curves of SGRBs in our sample.
The three SGRBs (050724, 131004A and 161004A) are shown
by red, green and blue stars, respectively.
The other sources are shown by gray curves.
}
\label{F2}
\end{figure*}

In a recent review, \citet{Berger2014} provides 67 \textit{Swift} SGRBs,
among which 31 events have X-Ray Telescope (XRT) rapid follow-up observations
($\la 100~\s$) and the adequate observational data in the early time
(100~\s~to 1000~\s)
\footnote{http://www.swift.ac.uk/xrtcurves/
\citep{Evans2007,Evans2009}
and http://www.astro.caltech.edu/grbox/grbox.php}.
In addition, we include 18 other \textit{Swift} SGRBs between
January 2013 to August 2017
\footnote{http://www.astro.caltech.edu/grbox/grbox.php}
with rapid and adequate XRT follow-up observations in our sample.
We have examined the total 49 SGRBs in our sample, and searched for
the bright X-ray flares satisfying the condition
``$F_{\p}> 3F$" \citep[e.g.,][]{Mu2016},
where $F_{\p}$ and $F$ are the peak flux and
the underlying continuum flux at the peak time, respectively.
Three SGRBs (050724, 131004A and 161004A) are found to have bright
X-ray flares, which are presented in Figure~\ref{F1}.
In this figure, the single power-law model (black solid line)
is adopted to fit the smooth continuum contribution
\citep[e.g.,][]{Bernardini2011,Margutti2011}.
All the X-ray light curves of the 49 SGRBs are presented in
Figure~\ref{F2}, where $z=0.5$ \citep{Berger2014} is adopted for
calculating the luminosity of the unknown redshift sources.
The three SGRBs (050724, 131004A and 161004A) are shown by red,
green and blue stars, respectively, and the other sources are shown by
gray curves.

An empirical function proposed by \citet{Norris2005} is used for fitting the
0.3 $-$ 10 \keV~(\Swift/XRT) X-ray flare light curve, i.e.,
\begin{equation}
F_{t}=A e^{2(\tau_{1}/\tau_{2})^{1/2}
- \frac{\tau_{1}}{t}-\frac{t}{\tau_{2}}} \ .
\end{equation}
The peak time of the flare is $t_{\p} = (\tau_{1}\tau_{2})^{1/2}$.
Thus, the peak count rate of the flare is $A=F_{\rm{max}}=F_{t_{\p}}$.
The width of the bright flare is measured between the two $1/e$
intensity points,
\begin{equation}\label{width}
\omega=\Delta t_{1/e}=\tau_{2}(1+4\mu)^{1/2} \ .
\end{equation}
where $\mu=(\tau_{1}/\tau_{2})^{1/2}$.
The asymmetry of the flare is
\begin{equation}\label{k}
k=\frac{t_{\d}-t_{\r}}{t_{\d}+t_{\r}}=(1+4\mu)^{-1/2} \ .
\end{equation}
where $t_{\d}=\omega(1+k)/2$ and $t_{\r}=\omega(1-k)/2$
are the decay and the rise time of the flare, respectively \citep{Bernardini2011}.
The fitting results of the X-ray flares are shown in Table~\ref{T1}.
The fitting procedure of the three SGRBs in our sample are shown
in Figure~\ref{F1}.
Results of the temporal evolution of the spectral photon index $\Gamma$
\footnote{\url{http://www.swift.ac.uk/burst_analyser/}}
are portrayed in Figure~\ref{F1}.

\begin{figure*}
\centering
\plotone{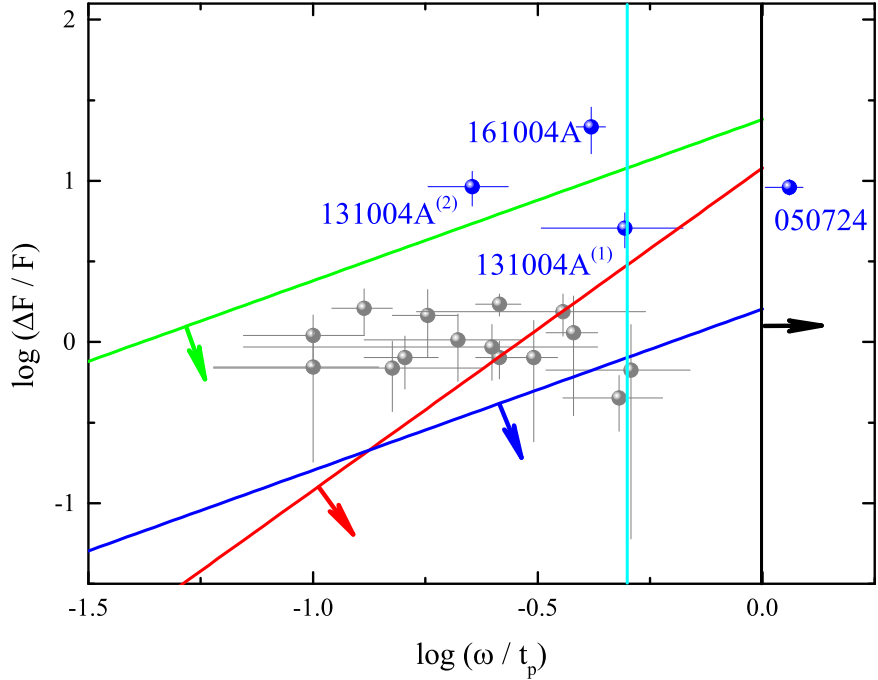}
\caption{
Relationship between the relative variability flux $\Delta F/F$
and the relative variability timescale $\omega /t_{\p}$ for
the four bright X-ray flares in our SGRB sample.
The bright flares in GRB 131004A and GRB 161004A are shown by
the blue circles, compared to the dim X-ray flare
candidates in short gamma-ray bursts (gray circles) from \citet{Margutti2011}.
The late-time bright X-ray flare detected in GRB 050724
($t_{\p}\sim 5\times 10^{4}~\s $) is also shown
by a blue circle.
The four theoretical lines are identical with those in
Figure~6 of \citet{Bernardini2011}, i.e., density fluctuations
on axis (blue line) and off-axis (red line), off-axis multiple regions
density fluctuations (green line), patchy shell model (black line),
see \citet{Ioka2005} for details.
In addition, the condition $\omega/t_{\p} < 0.5$ is shown by
the left region to the cyan vertical line.
}
\label{F3}
\end{figure*}

In order to investigate the physical origin of X-ray flares,
\citet{Ioka2005} proposed some limits on the timescale and
amplitude of variabilities to identify the sources of afterglow
variability, especially for $\omega < t_{\p}$.
It is a useful tool to trace the regions of allowance for the
timescale and amplitude of variabilities bumps or flares in
GRB afterglow on the basis of kinematic arguments.
In Figure~\ref{F3} we present our bright X-ray flares in
the ``Ioka plot" to judge the physical origin of the X-ray flares,
which is based on the relative variability flux
$\Delta F/F$ and the relative variability timescale $\omega /t_{\p}$.
The X-ray flare candidates in SGRBs from \citet{Margutti2011} are also
shown for the comparison.
Since X-ray flares generally follow a fast rise exponential decay profile
with $\omega \ll t_{\rm p} $, the criterion of $\omega < t_{\p}/2$
is adopted for the central engine-powered X-ray flares.
Similar to \citet{Ioka2005}, the fraction of cooling energy
$f_{c} \sim 1/2$ and $F/(\nu F_{\nu}) \sim 1$ are adopted for the X-ray band.
As shown in Figure~\ref{F3}, the bright X-ray flares from GRBs 131004A
and 161004A are well located in the upper left region
(above the green or red solid line and left to the cyan solid line),
which indicates that the flares are likely to be related to the
central engine reactivity.
The fitting parameters are shown in Table~\ref{T1}.

\begin{figure*}
\centering
\plotone{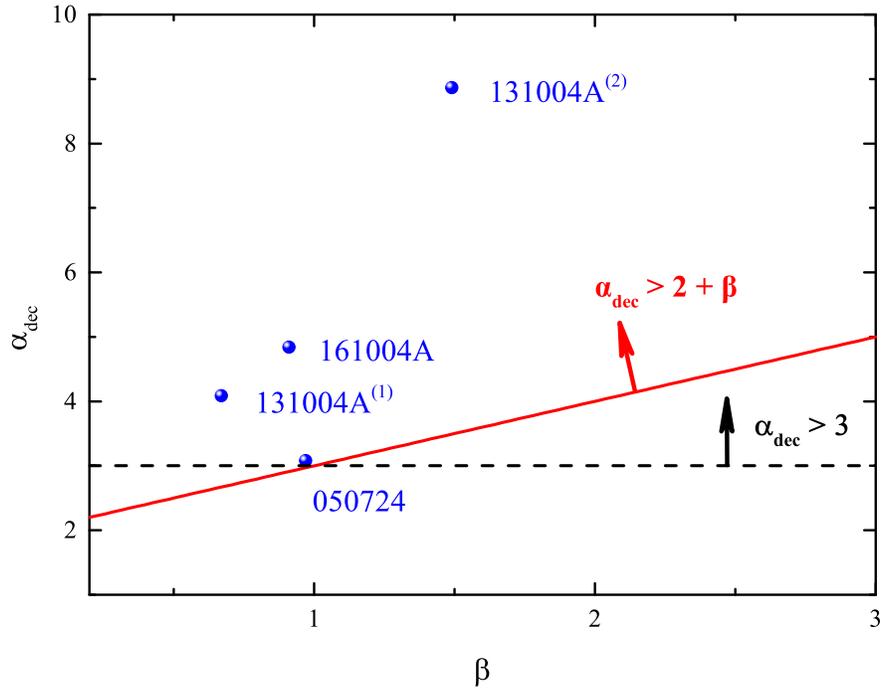}
\caption{
A comparison of the bright X-ray flares in our sample with
the criterion of internal origin ``$\alpha_{\dec} > 2+\beta$"
(the region above the red solid line)
and ``$\alpha_{\dec} > 3$" (the region above the black dashed line).
}
\label{F4}
\end{figure*}

On the other hand, by setting the zero time at the trigger time of GRB,
$\alpha_{\dec}= 2+\beta$ is the maximum decay index for the
external shock-powered flares
\citep[e.g.,][]{Kumar2000,Liang2006,Kobayashi2007},
where $\beta$ is the spectral index.
Then, $\alpha_{\dec}> 2+\beta$ is taken as another criterion
for the internal origin \citep{Ioka2005,Bernardini2011,Mu2016}.
An even simpler criterion takes the form $\alpha_{\dec}> 3$,
since $\beta$ is usually around unity.
The light curve index $\alpha_{\dec}$ in the decay phase
(from $t_{\p}$ to $t_{\p} + t_{\d}$) can be estimated as
\begin{equation}\label{lcdec}
\alpha_{\dec}=\frac{\log (e)}{\log [(t_{\p} + t_{\d})/t_{\p}]} \ ,
\end{equation}
The spectral index $\beta$ in the decay phase of the flare
is based on a power-law spectral model
\footnote{\url{http://www.swift.ac.uk/xrt spectra/addspec.php/}}.
The spectral analyses results are reported in Table~\ref{T1}.
In Figure~\ref{F4}, we present the flares
in the $\alpha_{\dec}$ - $\beta$ space.
It is seen from Figure~\ref{F4} that,
the flares from GRBs 131004A and 161004A
are located above the red solid line and the black dashed line, which implies
that they are likely to be of the internal origin.
Such a result is in good agreement with Figure~\ref{F3}.
We therefore argue that the flares from GRBs 131004A
and 161004A are related to the reactivity of the central engine.

\section{Accreted mass for X-ray flares}\label{sec3}

For the central engine-powered bright flares in GRBs 131004A and 161004A,
the isotropic energy of the flare $E_{\rm{X,iso}}$ in \Swift/XRT
energy range (0.3-10~\keV) is calculated by
\begin{equation}\label{Exiso}
E_{\rm{X,iso}}=\frac{4\pi D_{\rm{L}}^{2}S_{\rm{F}}}{1+z} \ ,
\end{equation}
where $D_{\rm{L}}$ is the luminosity distance of GRB.
The energy fluence in the energy range is
\begin{equation}\label{fluence}
S_{\rm{F}}=N\int^{10t_{\e}}_{0.1t_{\s}}A e^{2(\tau_{1}/\tau_{2})^{1/2}
- \frac{\tau_{1}}{t}-\frac{t}{\tau_{2}}} \ dt \ ,
\end{equation}
where $N$ is the factor of count-rate light curves converting into flux,
$t_{\s}$ and $t_{e}$ are the two $1/e$ intensity points,
i.e., the start and the end time of the flare.
We choose a sufficiently large time interval, i.e., from
$0.1t_{\s}$ to $10t_{\e}$, for the integration of Equation~(\ref{fluence}).
For GRB 161004A without redshift measurement,
we adopt the median redshift of the SGRB population $z=0.5$ \citep{Berger2014}.
The isotropic peak luminosity of the flare is
\begin{equation}
\label{LP}
L_{\p} = 4\pi D_{\rm{L}}^{2}AN\ .
\end{equation}
The isotropic energy and peak luminosity of X-ray flares are
reported in Table~\ref{T2}. The Blandford-Znajek (BZ) process
\citep{Blandford1977}
is a well-known mechanism to power GRBs and corresponding X-ray flares.
Based on a common assumption that the magnetic field $B$  is around
$10\%$ of its equipartition value \citep[e.g.,][]{Popham1999,Di2002},
the analytic BZ jet power
\footnote{
Since the neutrino annihilation mechanism may play an important role
only for extremely high mass accretion rates,
i.e., above the igniting accretion rate of an NDAF
\citep[e.g.,][]{Janiuk2007,Janiuk2013,Xue2013,Just2016,Lei2017},
here we focus on the BZ \citep{Blandford1977} mechanism, which is valid
for a wide range of mass accretion rates, even valid for sub-Eddington
accretion systems such as X-ray binaries and AGNs.}
can be expressed as \citep[e.g.,][]{Popham1999,Lee2000a,Lee2000b,
Di2002,Lei2017}
\begin{equation}
\label{PBZ}
\dot{E}_{\rm BZ} = \lambda(a_{*}) \times 10^{51}
\left( \frac{\dot M_{\rm in}}{M_{\sun}~{\rm s}^{-1}} \right)
\ {\rm erg~s}^{-1} \ ,
\end{equation}
where $\lambda(a_{*})$ is a function of $a_{*}$. Following
\citet{Lei2017} (but for a different ratio to the equipartition value),
we obtain $\lambda(a_{*})\simeq 1.8$, 3.1 and 4.4 for $a_{*} = 0.8$,
$0.9$, and $0.95$, respectively.
$\dot M_{\rm in} = M_{\rm frag}(1+z)/\omega$ is the average mass accretion
rate for the X-ray flares, with $M_{\rm frag}$ being the fragment mass.
In addition, we assume that the BZ jet power $\dot{E}_{\rm BZ}$
roughly equals
the isotropic luminosity $L_{\rm{X,iso}}$ of the flare \citep[e.g.,][]{Liu2018},
which means that the effects of the jet opening angle and the efficiency
from the jet power to the jet luminosity are comparable.
Then, the theoretical total energy of the X-ray flare can be expressed as
\begin{equation}
\label{EBZ}
E_{\rm{X,iso}} = L_{\rm{X,iso}} \frac{\omega}{1+z} \approx
\dot{E}_{\rm BZ} \frac{\omega}{1+z} = \lambda(a_{*})
\times 10^{51}~\frac{M_{\rm frag}}{M_{\sun}}~{\rm erg} \ .
\end{equation}

\begin{figure*}
\centering
\plotone{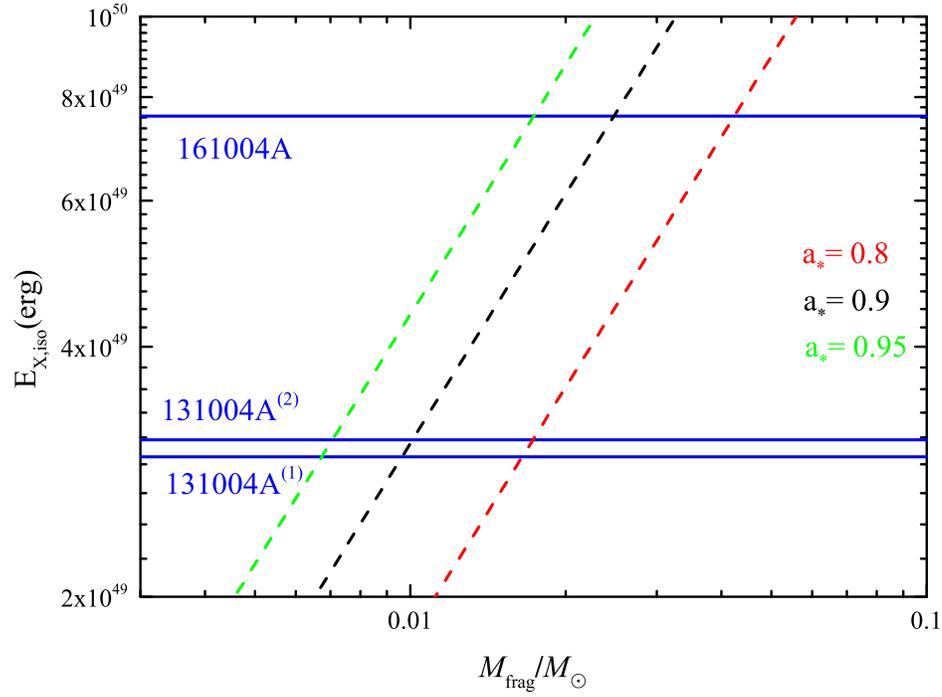}
\caption{
Relationship between the isotropic energy of the X-ray flares $E_{\rm X, iso}$
and the fragment mass $M_{\rm frag}$ based on the BZ mechanism.
The dashed lines represent the theoretical results according
to Equation~(\ref{EBZ}), where $a_{*} = 0.8$ (red), 0.9 (black)
and 0.95 (green). The isotropic energy $E_{\rm X, iso}$ for the flares
from GRBs 131004A and 161004A are shown by the blue solid lines.
}
\label{F5}
\end{figure*}

A comparison of the theoretical results with the observations is shown in
Figure~\ref{F5}, where the spin parameter is chosen as
$a_{*} = 0.8$ (red) , $0.9$ (black), $0.95$ (green).
It is seen from Figure~\ref{F5} that,
the BZ process can be responsible for the three flares in our sample.
For a typical spin parameter $a_{*} = 0.9$, the black dashed line
shows that the required fragment mass is
$\sim 0.01 M_{\sun}$ for each of the two flares of GRB 131004A,
and $\sim 0.025 M_{\sun}$ for the flare of GRB 161004A.

\section{A comparison of NS-NS and BH-NS mergers}\label{sec4}

A hyper-accreting stellar-mass BH or a millisecond magnetar
\citep[e.g.,][]{Dai1998,Zhang2001}
is usually invoked as the possible central engine of GRB.
This work will mainly focus on the hyper-accreting BH.
In the hyper-accreting stellar-mass BH,
a large amount of neutrinos can escape from the flow,
namely neutrino-dominated accretion flow
\citep[for a review of NDAF, see][]{Liu2017}.
Due to the high mass density of the NDAF, \citet{Liu2014} presented
the effects of self-gravity on the vertical structure and
the neutrino luminosity of the NDAF.
The instabilities of a hyperaccreting disk around
a BH may be responsible for the central
engine-powered bright X-ray flare.
The well-known disk instabilities, e.g., the thermal instability:
\begin{equation}
(d \ln Q^{+}/d \ln T)_{\Sigma} > (d \ln Q^{-}/d \ln T)_{\Sigma} \ ,
\end{equation}\
where $Q^{+}$ and $Q^{-}$ are the heating and cooling rates, respectively;
the viscous instability:
\begin{equation}
d \dot{m}/d \Sigma < 0 \ ,
\end{equation}\
and the gravitational instability: the Toomre parameter
\begin{equation}
Q_{\rm{T}} < 1 \ .
\end{equation}\

\citet{Di2002} suggested that the GRB disks are thermally stable
and viscously stable (see their Figure~4 for the stability analysis).
They also showed that $Q_{\rm{T}}$ decreases with increasing $r$,
so that the flows are most unstable in the outer regions.
However, only for the largest accretion rate
$\dot{M} \sim 10 M_{\sun} \rm{s^{-1}}$
and for $R \ga 50 R_{\rm{s}}$ does $Q_{\rm{T}} < 1$,
signifying gravitational instability \citep{Popham1999,Narayan2001,Di2002,Perna2006}.
The gravitational instability may be the most likely candidate
for the large-amplitude variability \citep{Perna2006}.
Moreover, \citet{Perna2006} discussed the physical conditions
in the outer parts of the hyperaccretion disk, and conclude that
gravitational instability, possibly followed by the actual
fragmentation of the disk, is the most likely candidate for
the large-amplitude variability of the central engine output
of both LGRBs and SGRBs.

The scenario is that, since the GRB disk is thermally and
viscously stable \citep{Popham1999,Narayan2001,Di2002}, the gravitational instability may be
the most likely candidate for the large-amplitude variability \citep{Perna2006}.
Once the disk is gravitationally unstable, two classes of behavior
are possible \citep[e.g.,][]{Perna2006,Liu2014}.
First, the disk may develop a quasi-steady spiral structure which can transfer
angular momentum outward and mass inward.
Second, if the local cooling of the hyperaccretion disk is rapid,
the disk may fragment into bound objects.
In such case, a small fraction of the disk materials may break away
from the main disk owing to
the gravitational instability, and therefore can account for the central
engine reactivity to power a bright X-ray flare.

After rapid accretion and/or merger of the initial fragments,
the fragment mass can be estimated as \citep{Takeuchi1996}
\begin{equation}
\label{fragment1}
M_{\rm frag} \simeq (\frac{H}{R})^{2}\alpha^{1/2}M_{\rm BH} \ ,
\end{equation}\
where $R$ is the distance from the accreting object, and
$\alpha$ is a dimensionless parameter characterizing the strength of viscosity
\citep{Shakura1973}.
Given $H \sim R$ and considering a range of the central compact object
masses between $1.5$ and $10 M_{\sun}$,
\citet{DallOsso2017} showed fragments with masses on
the order of $0.05 \sim 1 M_{\sun}$.
The geometrical thickness of a GRB disk is between the standard thin disk
and the advection-dominated accretion flow \citep[e.g.,][]{Narayan1994}.
Thus, in this paper, $H/R$ can be roughly estimated in the range
$0.1 \la H/R \la 0.2$
\footnote{If the extreme case $H \sim R$ is used to
estimate the fragment mass, and for a typical viscosity $\alpha = 0.02$ and
even a low-mass BH $M_{\rm BH} = 3M_{\sun}$, $M_{\rm frag}$ is over
$0.4 M_{\sun}$ according to Equation~(\ref{fragment1}). Such a value is even
beyond the total disk mass before fragment under the NS-NS merger,
and therefore likely unphysical.}
according to previous calculations on NDAF solutions
\citep[e.g.,][]{Popham1999,Kawanaka2013}, particularly for the inner disk.
In addition, the viscosity parameter is chosen as $\alpha = 0.02$
according to MHD simulation results \citep{Hirose2009}.

\begin{figure*}
\centering
\plotone{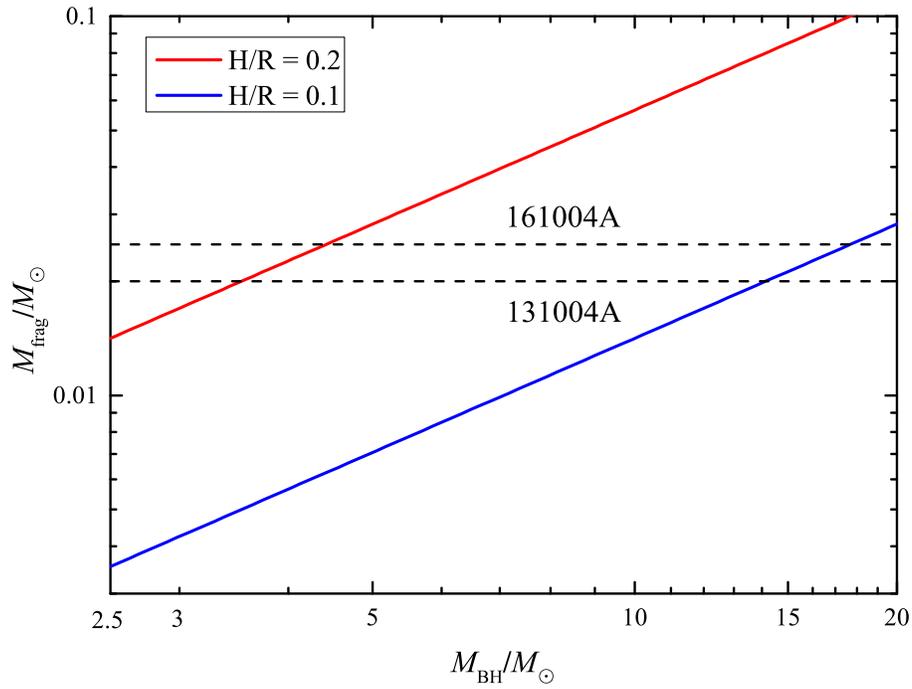}
\caption{
Relationship between the fragment mass $M_{\rm frag}$ and the
central BH mass $M_{\rm BH}$, where the viscosity parameter
$\alpha = 0.02$. The blue and red solid lines represent the theoretical
results according to Equation~(\ref{fragment1})
for $H/R =0.1$ and $0.2$, respectively.
The required fragment masses of the X-ray flares from GRBs 131004A and 161004A
are shown by the black dashed lines, which are estimated by
Equation~(\ref{EBZ}) with a typical spin parameter $a_{*} = 0.9$.
}
\label{F6}
\end{figure*}

The relationship between the fragment mass and BH mass
is presented in Figure~\ref{F6}.
According to previous simulations \citep{Ruffert1997,Ruffert1999},
the NS-NS merger may result in a BH around $2.5 \sim 3 M_{\sun}$,
surrounded by a disk of around $0.1 \sim 0.2 M_{\sun}$.
In Figure~\ref{F6}, we consider the BH masses in the range
$2.5 \sim 20 M_{\sun}$.
The solid lines represent the theoretical $M_{\rm frag}$ for $H/R = 0.1$
(blue) and $0.2$ (red). As shown by the red line, the upper critical
$M_{\rm frag}$ at $M_{\rm BH} = 3M_{\sun}$ is $\sim 0.017 M_{\sun}$,
which is a reasonable fraction (around 10\%) of the upper limit of
the total disk mass ($0.1 \sim 0.2 M_{\sun}$) during the NS-NS merger,
according to previous simulations \citep[e.g.,][]{Ruffert1997,Dietrich2015}.
For the BH-NS merger, the upper critical $M_{\rm frag}$ is
$\sim 0.057 M_{\sun}$, also around 10\% of the upper limit of
the total disk mass ($\sim 0.5 M_{\sun}$
\citealp[e.g.,][]{Kluzniak1998,Janka1999}).
On the other hand, it is known that the typical energy ratio of the X-ray
flare to the prompt gamma-ray emission is around 10\%
\citep[e.g.,][]{Chincarini2010,Yi2016}.
Thus, the theoretical thresholds for $M_{\rm frag}$ (the two solid lines
in Figure~\ref{F6}) well agree with the observations and therefore
may be reasonable.

For the two SGRBs, as shown in Figure~\ref{F5},
the fragment mass is around $0.02 M_{\sun}$ for the sum of two flares
of GRB 131004A, and $0.025 M_{\sun}$ for the flare of GRB 161004A,
as shown by the two dashed lines in Figure~\ref{F6}.
It is seen that these two values are just beyond the upper critical
$M_{\rm frag}$ under the NS-NS merger (at $M_{\rm BH} \sim 3M_{\sun}$).
On the contrary, the BH-NS merger can provide a larger fragment mass,
i.e. up to $\sim 0.057 M_{\sun}$,
so it can be responsible for the
bright X-ray flares.
It is also seen from Figure~\ref{F6} that, the two dashed lines
(the parts between the two solid lines) indicate that the central BH masses
of GRBs 131004A and 161004A are likely in the range $3.5 \sim 15 M_{\sun}$,
corresponding to the BH-NS merger rather than the NS-NS merger.

However, we should stress that, due to the large uncertainty for the
power and efficiency of the BZ process, the required fragment mass
$M_{\rm frag}$ for the X-ray flare is quite uncertain.
The NS-NS merger may account for bright X-ray flares if the efficiency
from the jet power to the jet luminosity is higher, or the BH spin is faster.
In addition, even for the normal efficiency and spin parameter,
the NS-NS merger can also be responsible for the flares with relatively
low energy. Nevertheless, according
to Equation~(\ref{fragment1}), under the same parameters, the BH-NS
merger will generally have a larger $M_{\rm frag}$ than the NS-NS merger.
We therefore argue that the BH-NS merger is more likely to be the
central engine of GRBs with bright X-ray flares.

\section{Conclusions and discussion}\label{sec5}

In this work, we have studied a sample of 49 SGRBs with
rapid \textit{Swift}/XRT follow-up observations,
and the adequate observational data in the early time.
We have found that three bursts have four bright X-ray flares,
among which three bright X-ray flares are probably related to the
central engine reactivity.
Since the GRB disk is thermally and
viscously stable \citep{Popham1999,Narayan2001,Di2002},
the gravitational instability, possibly followed by fragmentation,
may be the most likely candidate for the bright X-ray flares \citep{Perna2006}.
Moreover, a comparison of the analytic BZ jet power with the observations
is presented in this paper. We have found that, based on the BZ process,
the fragment mass is sufficient for the required mass accretion for the
bright flares in our sample. In our scenario, GRBs 131004A
and 161004A are likely related to a massive disk around a BH,
and therefore the BH-NS merger is more preferred than
the NS-NS merger.

We should stress that our study is based on the assumption that the central
engine is a BH hyper-accretion, formed by the BH-NS merger or the NS-NS merger.
Apart from such a mechanism,
a magnetar produced by the NS-NS merger can also be responsible for the
central engine of the X-ray flares in SGRBs \citep[e.g.,][]{Dai2006}.
It is well-known that the GW detectors advanced LIGO and Virgo have detected
the GW radiation from the BH-BH and NS-NS mergers. We can expect
that the GW radiation of BH-NS merger will also be detected
in near future. Thus, our suggested link between the central
engine-powered bright X-ray flare and the BH-NS merger can be checked.

The local event rate density for SGRBs varies slightly for different
merger delay models: Gaussian, lognormal, and power-law delay models \citep{Sun2015}.
It is known that the BH-NS merger rate is significantly lower
than the NS-NS one. \citet{Voss2003} shows that in our Galaxy the former is
roughly 40\% of the latter.
In addition, in the present work we only consider the bright X-ray flares
and ignore the possibility of dim flares originating from the BH-NS merger.
Moreover, the disk mass during the BH-NS merger is sensitive to
the mass ratio of the two compact objects. In other words,
a BH-NS merger is likely a necessary but not a sufficient condition
to have a massive disk.
Thus, the percentage of the central engine-powered bright flares
in our sample, $2/49 \simeq 4\%$,
is likely a reasonable value according to our suggested link.

\acknowledgments

We thank the referee for beneficial suggestions that improved the manuscript.
We acknowledge the use of the public data from the \textit{Swift} data
archive, and the UK \textit{Swift} Science Data Center.
This work was supported by the National Basic Research Program of China
(973 Program) under grants 2014CB845800,
and the National Natural Science Foundation of China under grants
11573023, 11673062, 11473022, 11333004, 11503011, 11773007 and 11403005.
W.-M. Gu is supported by the CAS Open Research Program of Key Laboratory
for the Structure and Evolution of Celestial Objects under grant OP201503.
J. Mao is supported by the Hundred Talent Program, the Major Program
of the Chinese Academy of Sciences (KJZD-EW-M06),
and the Oversea Talent Program of Yunnan Province.

\clearpage

\clearpage

\clearpage

\clearpage

\clearpage

\clearpage

\clearpage

\begin{deluxetable}{lccccccccccc}
\tablecaption{Fitting results of the flares in our sample.}
\centering
\tablewidth{0pt}
\tabletypesize{\scriptsize}
\tablecolumns{12}
\tablenum{1}
\tablehead{
\colhead{$\rm{GRB}$}
&\colhead{$T_{90}$\tablenotemark{a}}
&\colhead{$A$}
&\colhead{$\tau_{1}$}
&\colhead{$\tau_{2}$}
&\colhead{$k$}
&\colhead{$\omega$}
&\colhead{$\omega/t_{\rm{p}}$}
&\colhead{$\Delta F/F$}
&\colhead{$\alpha_{\dec}$}
&\colhead{$\beta$}
&\colhead{$N$\tablenotemark{b}}
\\
\colhead{ }
&\colhead{(\s)}
&\colhead{$(\rm{counts~s^{-1}})$}
&\colhead{$(\rm{10^{4}~s})$}
&\colhead{$(\s)$}
&\colhead{ }
&\colhead{$(\s)$}
&\colhead{ }
&\colhead{ }
&\colhead{ }
&\colhead{ }
&\colhead{$(\rm{erg~cm^{-2}})$}
}
\startdata
050724\tablenotemark{c}	&3	&	0.035 	$\pm$	0.003 	&	21.6	$\pm$	4.62	&	$(1.2 \pm 0.2)\times10^{4}$ 	&	0.12 	&	$5.09\times10^{4}$ 	&	1.15 	$\pm$	0.19 	&	9.12 	$\pm$	3.40 	&	3.08 	&	0.97	&	-	\\
131004A$^{(1)}\tablenotemark{d,f}$	&	1.54	&	2.45 	$\pm$	0.53 	&	0.82 	$\pm$	0.41 	&	29.77 	$\pm$	14.71 	&	0.06 	&	244.7 	&	0.49 	$\pm$	0.17 	&	5.08 	$\pm$	1.24 	&	4.09 	&	0.67	&	4.02E-11	\\
131004A$^{(2)}\tablenotemark{e}$	&	1.54	&	3.50 	$\pm$	0.83 	&	6.26 	$\pm$	1.82 	&	10.16 	$\pm$	2.87 	&	0.10 	&	180.2 	&	0.23 	$\pm$	0.05 	&	9.22 	$\pm$	2.24 	&	8.86 	&	1.49	&	4.02E-11	\\
161004A\tablenotemark{g}	&	1.3-3	&	21.8 	$\pm$	3.7 	&	0.72 	$\pm$	0.08 	&	13.29 	$\pm$	1.33 	&	0.24 	&	129.1 	&	0.42 	$\pm$	0.03 	&	21.67 	$\pm$	6.93 	&	4.84 	&	0.91	&	4.64E-11	\\
\enddata
\tablenotetext{a}{duration in the 15-350 \keV~band.}
\tablenotetext{b}{$N$ is the factor of count-rate light curves converting
into flux, taking GRB 161004A as an example,
$1~\rm{count} = 4.64\times10^{-11} \rm{erg~cm^{-2}}$.}
\tablenotetext{c}{See \citet{Barthelmy2005,Campana2006,Grupe2006} for GRB 050724.}
\tablenotetext{d}{the first bright X-ray flare in GRB 131004A.}
\tablenotetext{e}{the second bright X-ray flare in GRB 131004A.}
\tablenotetext{f}{\citet{Stamatikos2013} reported that the spectrum of
GRB 131004A appears at the softer end of short bursts,
which have an average \\ power-law spectral index of 1.2.
The spectral lag for the burst
is 0.130 \s~ $\pm$ 0.020 \s~(for the 50-100 \keV~and 15-25 \keV~bands),
which \\ is significantly longer than the regular SGRBs.}
\tablenotetext{g}{\citet{Barthelmy2016} showed that $T_{90}$ of GRB 161004A
varies from $\sim$ 1.3 \s to $\sim$ 3 \s when using light
curves with different bin sizes.}
\label{T1}
\end{deluxetable}

\clearpage

\begin{deluxetable}{lcccc}
\tablecaption{Physical parameters of the flares.}
\centering
\tablewidth{0pt}
\tabletypesize{\small}
\tablecolumns{5}
\tablenum{2}
\tablehead{
\colhead{$\rm{GRB}$ }
&\colhead{$z$}
&\colhead{$\omega_{\rm{rest}}$\tablenotemark{a}}
&\colhead{$E_{\rm {X,iso}}$\tablenotemark{b}}
&\colhead{$L_{\p}$\tablenotemark{c}}\\
\colhead{ }
&\colhead{ }
&\colhead{(\rm{s})}
&\colhead{$(10^{49}~\erg)$}
&\colhead{$(10^{47}~\rm{erg~s^{-1}})$}
}
\startdata
050724	&	0.257	&	$4.049\times10^{4}$	&	-	&	-	\\
\hline
131004A$^{(1)}$	&	0.717	&	142.5	&	2.95	&	$2.32	\pm	0.51$	\\
131004A$^{(2)}$	&	0.717	&	104.9	&	3.09	&	$3.32	\pm	0.79$	\\
161004A	&	-	&	86.09	&	7.59	&	$9.92	\pm	1.67$	\\
\enddata
\tablenotetext{a}{the width of the flare in the rest frame.}
\tablenotetext{b}{the isotropic energy of the X-ray flare. $z =0.5$ for GRB 161004A.}
\tablenotetext{c}{the isotropic peak luminosity of the flare.}
\label{T2}
\end{deluxetable}

\end{document}